\documentclass[12pt,preprint]{aastex}
\newcommand{\etal}{\hbox{ et~al.}}

\begin{document}



\title {Observations of the Lensed Quasar Q2237+0305 with CanariCam at GTC}


\author{H. Vives-Arias$^1$, J. A. Mu\~noz$^{1,2}$, C. S. Kochanek$^{3}$, E. Mediavilla$^{4,5}$, and
J. Jim\'enez-Vicente$^{6,7}$}
\bigskip

\affil{$^{1}$Departamento de Astronom\'{\i}a y Astrof\'{\i}sica, Universidad
       de Valencia, E-46100 Burjassot, Valencia, Spain}
\affil{$^{2}$Observatorio Astron\'omico, Universidad de Valencia, E-46980 Paterna, Valencia, Spain}        
\affil{$^{3}$Department of Astronomy, The Ohio State University, 140 West 18th Avenue, Columbus, OH 43210, USA}
\affil{$^{4}$Instituto de Astrof\'{\i}sica de Canarias, E-38200 La Laguna, Santa Cruz de Tenerife, Spain}
\affil{$^{5}$Departamento de Astrof\'{\i}sica, Universidad de La Laguna, E-38200 La Laguna, Santa Cruz de Tenerife, Spain}
\affil{$^6$Departamento de F\'{\i}sica Te\'orica y del Cosmos, Universidad de Granada, Campus de Fuentenueva, E-18071
Granada, Spain}
\affil{$^7$Instituto Carlos I de F\'{\i}sica Te\'orica y Computacional, Universidad de Granada, E-18071 Granada, Spain}   


\begin{abstract}

We present new mid-IR observations of the quadruply lensed quasar Q2237+0305 taken with CanariCam on the Gran Telescopio Canarias. Mid-IR emission by hot dust, unlike the optical and near-IR emission from the accretion disk, is unaffected by the interstellar medium (extinction/scattering) or stellar microlensing. We compare these ``true" ratios to the (stellar) microlensed flux ratios observed in the optical/near-IR to constrain the structure of the quasar accretion disk. We find a half-light radius of $R_{1/2}=3.4_{-2.1}^{+5.3}\sqrt{\langle M \rangle/0.3\,\rm{M_{\sun}}}$ light-days at $\lambda_{rest}=1736$ \AA,
and an exponent for the temperature profile $R \propto \lambda^{p}$ of $p=0.79\pm0.55$, where 
$p=4/3$ for a standard thin-disk model.
If we assume that the differences in the mid-IR flux ratios measured over the years are due to microlensing variability, we find a lower limit for the size of the mid-IR-emitting region of $R_{1/2} \gtrsim 200\,\sqrt{\langle M \rangle/0.3\,\rm{M_{\sun}}}$ light-days.
We also test for the presence of substructure/satellites by comparing the observed mid-IR flux ratios with those predicted from
smooth lens models.
We can explain the differences if the surface density fraction in satellites near the lensed images is $\alpha = 0.033_{-0.019}^{+0.046}$ for 
a singular isothermal ellipsoid plus external shear mass model or $\alpha = 0.013_{-0.008}^{+0.019}$ for a mass
model combining ellipsoidal NFW and de Vaucouleurs profiles in an external shear. 

\end{abstract}

\keywords{ gravitational lensing --- dark matter --- accretion, accretion disks --- quasars: individual (Q2237+0305)}

\section{Introduction}\label{sec1}

Gravitational lenses are a powerful tool for many astrophysical and cosmological studies
(e.g. see the review by Kochanek 2006). In particular, multiply imaged quasars allow us to probe many properties of source quasars and the mass distribution and interstellar medium (ISM) of the lens galaxies that are hard to characterize otherwise. The flux ratios of the images, one of their most easily measured properties, are controlled not only by the smooth gravitational potential of the lens but also by perturbations produced by stars (microlensing) and satellites/cold dark matter (CDM) substructure (millilensing), as well as propagation effects in the lens (scattering/extinction). As a result, smooth lens models almost always fail to fit image flux ratios and thus are rarely used as model constraints. 

Optical and near-IR flux ratios can be altered by differential extinction between the lensed images (e.g. Nadeau et al. 1991, Falco et al. 1999, Mu\~noz et al. 2004, El{\'{\i}}asd{\'o}ttir et al. 2006). While we can try to correct for this by fitting extinction models, microlensing by the stars in the lens galaxy also produces color changes between images that can mimic extinction (e.g. Poindexter et al. 2008, Mosquera et al. 2009, Mu\~noz et al. 2011), so the two effects cannot be fully separated. Radio lenses generally avoid this problem (see Kochanek \& Dalal 2004), although there are clear examples of images that are scatter broadened (e.g. Sykes et al. 1998). Unfortunately, radio lenses are also a minority of lenses and in many cases lack the ancillary information needed to make them useful astrophysical probes (redshifts and well-studied lens galaxies). Mid-IR wavelengths are almost ideal for measuring the intrinsic flux ratios of lensed images. They are too long (short) to be affected by extinction (electrons), thereby eliminating the ISM as a factor affecting the flux ratios.
Because the mid-IR emission is dominated by hot dust, which is destroyed if too close to the quasar (e.g. Barvainis 1987), the mid-IR emission
regions should also be large enough to be little affected by microlensing.

This means that the deviations of the mid-IR flux ratios from models primarily probe the mean gravitational potential of
the lens and the effects of
substructure. This is astrophysically important because the amount of substructure in
CDM halos is an open question. Simulations suggest that 10\% of the mass remains in satellites, with the fraction dropping closer to 1\% in the inner regions as tidal effects destroy the satellites (e.g. Zentner \& Bullock 2003),
in contradiction with the observations of the Milky Way halo (Klypin et al. 1999, Moore et al. 1999). 
While this discrepancy has been reduced over the past decade with the discovery of many more faint Milky Way satellites (Belokurov et al. 2006; Zucker et al. 2006; Koposov et al. 2015), and new estimates of our halo mass predict fewer high-mass satellites (Wang et al. 2012; Kafle et al. 2014), it still exists.
Gravitational lensing is one of the only
means of detecting dark substructures, and results from studying anomalous flux ratios in radio lenses 
(Dalal \& Kochanek 2002, Kochanek \& Dalal 2004) and extended emission from host galaxies (Vegetti et al. 2012) suggest that the missing satellites are
present. However, mid-IR observations, as one of the best probes for the effects of substructure, are available for only six lenses (Agol et al.
2000, 2001, 2009, Chiba et al. 2005, MacLeod et al. 2009, 2013, Minezaki et al. 2009).

Q2237+0305 (Huchra et al. 1985) is a gravitational lens system where a relatively nearby spiral galaxy ($z_L=0.039$) creates four images of a much more distant quasar ($z_S=1.695$). The closeness of the lens galaxy to the observer makes the light paths of the multiple images go through the dense galactic bulge and leads to a high effective transverse velocity between the lens, the source, and the observer. This leads to short time scales for stellar microlensing variability, which has now been observed for $\sim$30 years (e.g. Corrigan et al. 1991, Webster et al. 1991, Wo\'{z}niak et al. 2000). Furthermore, because the light paths for the different images are so similar, the time delays between intrinsic brightness variations from the quasar are less than 1 day (Dai et al. 2003). Since quasars have little variability power on such short time scales, there is no need to correct for the delays in this system when interpreting single epochs of data.

In Section 2 we describe the GTC observations of Q2237+0305. In Section 3 we discuss the mid-IR flux ratios between the four images of the lensed quasar and compare them with previous observations and the predictions from lens models. In Section 4 we use these new estimates to recalculate the wavelength-dependent size of the quasar accretion disk. Section 5 estimates the abundance of substructure in the lens galaxy, and we summarize all these results in Section 6.

\section{Observations and Data analysis}\label{sec2}\

The mid-IR observations of Q2237+0305 were performed using the CanariCam imager on the Gran Telescopio Canarias (GTC), located at the Roque de los Muchachos Observatory, La Palma (Spain), in 2012 July and 2013 September. CanariCam has a field of view of 25\farcs6$\times$19\farcs2 with a spatial scale of 0\farcs08 pixel$^{-1}$.
For the filters we use, the resolution is diffraction limited by the 10.4 m primary mirror of GTC. For all observations, we set a chopping position angle of 53\degr, a nodding position angle of $-127$\degr, and a throw of $10\arcsec$ for both motions. The mid-IR standard stars HD~220009 and HD220954 were observed for each epoch of observation to be used as point-spread function (PSF)
templates for the data reduction.

\begin{deluxetable}{lcccl}
\tabletypesize{\footnotesize}
\tablewidth{0pt}
\tablecaption{Log of Q2237+0305 Observations with CanariCam}
\tablehead{
\colhead{Date} & \colhead{Filter} & \colhead{Readout Mode} & \colhead{Exposure (s)} & \colhead{Notes}}
\startdata
2012 Jun 6 & N-10.36 & S1R1\_CR & 1001.9 & Detected, non-Gaussian noise \\
2012 Jul 10 & N-10.36 & S1R1\_CR & $3\times675.3$ & Detected, non-Gaussian noise \\
2013 Sep 4 & Si5 & S1R3 & $3\times595.7$ & Nondetection \\
2013 Sep 18 & Si5 & S1R3 & $3\times1853.3$ & Detected, low S/N in third image \\
2013 Sep 19 & Si5 & S1R3 & $2\times1522.4$ & Detected \\
\enddata
\label{obs}
\end{deluxetable}

A test image was obtained on 2012 June 6 with an on-source exposure time of 1001.9 s using the S1R1\_CR readout mode, a chopping frequency of 2.05 Hz, and the N-10.36 filter ($\lambda_c = 10.36\, \mu$m, $\Delta\lambda = 5.2\, \mu$m). Since the object was successfully detected, three more images with on-source exposure times of 675.3 s each and a chopping frequency of 2.01 Hz were obtained on 2012 July 30. The S1R1\_CR mode, however, introduced a non-Gaussian horizontal noise pattern in the images that makes it difficult to accurately measure the fluxes of targets with low signal-to-noise ratios (S/Ns). The horizontal bands could be removed in the area of interest by selecting a range of columns with the same noise pattern as the region of the image where the target is located, averaging them and subtracting the pattern from the whole image. However, since we are interested in measuring flux ratios between images at different locations on the image, it is better to avoid this kind of noise altogether.

For the next set of observations, we switched to the newly available S1R3 readout mode, in which the noise pattern has a more Gaussian structure and the same properties along lines and columns. Unfortunately, the new mode also uses longer frame times, leading to high backgrounds that more easily saturate the detector. As a result, the N filter is not recommended for use in this mode unless the precipitable water vapor (PWV) is below 3 mm (which happens only around 2\% of the observing time). For this reason, we switched to the narrower Si5 filter ($\lambda_c = 11.6\, \mu$m, $\Delta\lambda = 0.9\, \mu$m). On 2013 September 4, three images were obtained with on-source exposure times of 595.7 s each and a chopping frequency of 2.07 Hz, but the target was not detected due to the smaller width and lower transmission of the Si5 filter. We then increased the exposure times for a last set of observations using the same configuration to obtain three images on September 18 and two on September 19 with total on-source exposure times of 3$\times$1853.3 and 2$\times$1522.4 s, respectively. A summary of all our observations can be found in Table~\ref {obs}.

\begin{figure}
\figurenum{1}
\epsscale{0.5}
\plotone{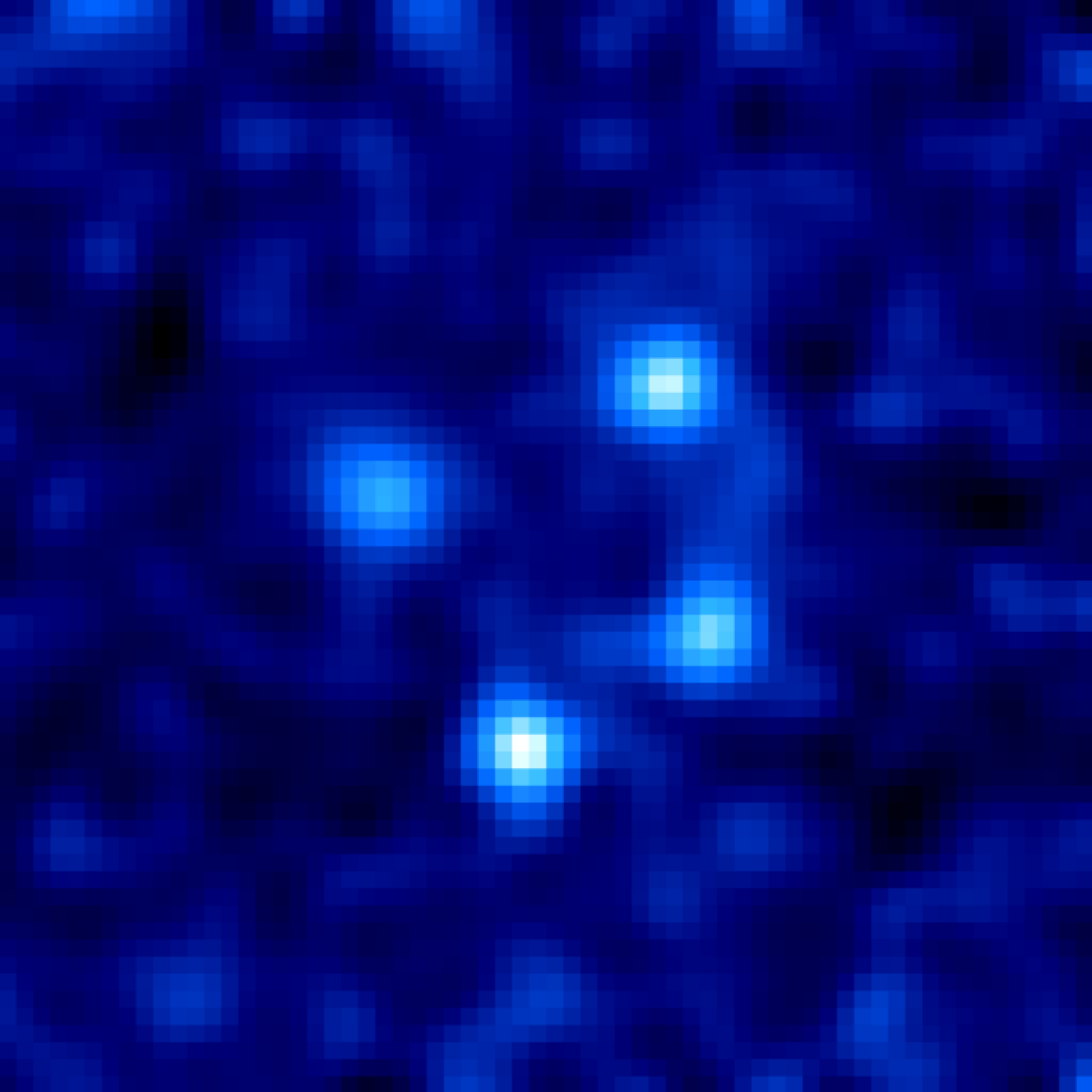} 
\caption{Quadruple lens system Q2237+0305 at $11.6\, \mu$m using data taken with CanariCam at GTC on 2013 September 18 and 19 (UT). The pixel scale is $0\farcs08$ pixel$^{-1}$, and the image subtends $5\farcs12$. North is up, east is left, and the quasar images are, starting from top right and moving clockwise, B, D, A, and C, respectively. This image has been smoothed with a Gaussian kernel of $\sigma = 0\farcs12$ in order to improve the contrast relative to the noise.}
\label{canaricam}
\end{figure}

The data were reduced by first aligning the images from each night of observation and then separately combining the images from the 2012 and 2013 observations. To determine the offset between the individual images for the alignment, we performed PSF fitting relative to the known locations of the quasar images from the \textit{Hubble Space Telescope} (HST) observations available on the CASTLES Web site\footnote{http://www.cfa.harvard.edu/glensdata/}. The final combined image (Figure~\ref{canaricam}) used only the 2013 observations due to their better instrumental conditions, excluding the third image from September 18, which had a very poor S/N due to a significant rise in the PWV. Experiments including this third image and/or the shorter exposures from September 4 did not lead to improved results. The raw FWHM of 
the quasar images is $0\farcs24$, which we have smoothed with a Gaussian kernel of $\sigma = 0\farcs12$ for
Figure~\ref{canaricam}.

\section{Flux ratios}\label{sec3}\

The flux ratios of the lensed images were obtained using PSF-fitting photometry from the combined 2013 image.
The scatter between the results from applying this same procedure to the individual noncombined images was used to estimate the errors. The final flux ratios are B/A $=0.99\pm0.03$, C/A $=0.69\pm0.10$, and D/A $=0.84\pm0.13$. The 2012 observations give flux ratios that are consistent with those from 2013 but with larger error bars (see Table~\ref {ratios}).

As shown in Table~\ref {ratios}, our measured B/A and C/A flux ratios differ significantly from the ones obtained by Minezaki et al. (2009) at $\lambda = 11.67 \mu$m in 2005 October with Subaru. Interestingly, they are compatible at the 1$\sigma$ level with the ratios measured by Agol et al. (2009) at $\lambda = 8.0 \mu$m in 2005 November with the \textit{Spitzer Space Telescope}. Previous observations at $\lambda = 11.7 \mu$m from 2000 November with the Long Wavelength Spectrometer on Keck by Agol et al. (2001) yielded flux ratios that are inconsistent with the ones measured by Minezaki et al. (2009) but much closer to our estimates. Only the D/A flux ratio shows a significant difference. 

The flux ratios measured by Minezaki et al. (2009) are in agreement with the prediction yielded by a simple singular isothermal ellipsoid (SIE) plus external shear ($\gamma$) model without the need for any additional structure when taking only the HST image positions as constraints. On the other hand, the Trott et al. (2010) mass model, consisting of a galactic bulge, bar, and disk combined with a dark matter halo fitted to the image positions and the observed kinematics of the galaxy but not the flux ratios, predicts fluxes that are closer to the ones observed by Agol et al. (2000). We discuss the consequences of these flux ratio differences further in Section~\ref{sec4}.

To compare the mid-IR flux ratios to the predictions of smooth and relatively simple mass models, we used the Gravlens/Lensmodel code \citep{keeton01,keeton11} to fit the image positions. In addition to the simple SIE + $\gamma$ model, we also considered a more elaborate model with a Navarro--Frenk--White (NFW) dark matter halo and a de Vaucouleurs profile for the bulge and bar of the lens galaxy. To reduce the number of free parameters for this second model, we constrained the ellipticity ($e$) and position angle ($\theta_e$) of the de Vaucouleurs profile to agree with estimates from fitting the HST images. A more relaxed condition was set for the effective radius ($R_e$), since this can vary among different filters. The parameters for the break radius $r_s$ and the surface density at the break radius $\kappa_s$ of the NFW profile are constrained to be close to those expected for a $\sim10^{12}\, \rm{M_{\odot}}$ halo (Trott et al. 2010), and we favored models with a small ellipticity to avoid unphysical solutions. 
It can be seen in Table~\ref {ratios} that the flux ratio predictions from our two models are very similar, and different from the ones in Trott et al. (2010).  

\begin{deluxetable}{cccccc}
\tabletypesize{\footnotesize}
\tablewidth{0pt}
\tablecaption{Mid-IR Flux Ratios for Q2237+0305}
\tablehead{
\colhead{Reference\tablenotemark{\dagger}} & \colhead{Date} & \colhead{Wavelength / Model} & \multicolumn{3}{c}{Flux Ratio}  \\
 & & & B/A & C/A & D/A }
\startdata
This work & 2013 Sep 18, 19 & 11.6 $\mu$m & $0.99\pm0.03$ & $0.69\pm0.10$ & $0.84\pm0.13$ \\
This work & 2012 Jul 10 & 10.36 $\mu$m & $0.96\pm0.11$ & $0.57\pm0.10$ & $1.04\pm0.21$ \\
1 & 2005 Nov 17 & 8.0 $\mu$m & $0.93\pm0.07$ & $0.59\pm0.04$ & $0.84\pm0.08$ \\
2 & 2005 Oct 11, 12 & 11.7 $\mu$m & $0.84\pm0.05$ & $0.46\pm0.02$ & $0.87\pm0.05$ \\
3 & 2000 Jul 11 & 11.7 $\mu$m & $1.11\pm0.09$ & $0.72\pm0.07$ & $1.17\pm0.09$ \\
4 & 1999 Jul 28, Sep 24 & 8.9 and 11.7 $\mu$m & $1.11\pm0.11$ & $0.59\pm0.09$ & $1.00\pm0.10$ \\
4 & 1999 Sep 24 & 11.7 $\mu$m & $0.91\pm0.30$ & $0.41\pm0.21$ & $0.66\pm0.27$ \\
4 & 1999 Sep 24 & 8.9 $\mu$m & $0.88\pm0.28$ & $0.51\pm0.22$ & $1.05\pm0.31$ \\
4 & 1999 Jul 28 & 11.7 $\mu$m & $1.07\pm0.25$ & $0.61\pm0.16$ & $1.09\pm0.25$ \\
4 & 1999 Jul 28 & 8.9 $\mu$m & $1.42\pm0.33$ & $0.66\pm0.20$ & $1.09\pm0.27$ \\
Average & 2013, 2005, 2000, and 1999 & 8.0 -- 11.6 $\mu$m & $0.97\pm0.03$ & $0.51\pm0.02$ & $0.92\pm0.04$ \\
\hline
This work & & SIE + $\gamma$ & 0.89 & 0.45 & 0.82 \\
This work & & NFW + de Vaucouleurs & 0.88 & 0.48 & 0.87 \\
5 & & Bulge + disk + halo + bar & 1.02 & 0.56 & 1.19 \\
\enddata
\tablenotetext{\dagger}{References: (1) Agol et al. 2009; (2) Minezaki et al. 2009; (3) Agol et al. 2001; (4) Agol et al. 2000; (5) Trott et al. 2010.}
\label{ratios}
\end{deluxetable}

\section{Accretion disk size estimation}\label{sec4}

Since the magnitude of the microlensing of the quasar images depends on the projected size of the source compared to the average Einstein radius of the microlenses, microlensing can be used to determine the size of the accretion disk, or other emission regions. The temperature of the disk is also expected to increase radially toward the center, so observations in different optical bands should give different results because shorter wavelengths correspond to smaller, more central, higher-temperature regions of the disk. These chromatic effects can be used to determine the scaling of the disk temperature with radius.

\cite{paper2237}, using a Bayesian analysis of six epochs of observations of Q2237+0305 in five narrowband filters over the wavelength range 4670-8130 \AA, combined with the Minezaki et al. (2009) mid-IR observations as an estimate of the intrinsic flux ratios, estimated two parameters of the disk, its half-light radius ($R_{1/2}$) and the logarithmic scaling slope ($p$) of its temperature profile $T \propto R^{-1/p}$. \cite{paper2237} found, as expected from earlier studies (Mortonson et al. 2005), that the half-light radius ($R_{1/2}$) estimates are independent of the surface brightness profile. Here we recalculate these two disk parameters using our new mid-IR observations. We assumed a standard thin-disk model, $I(R) \propto \left( \exp\left[(R/r_s)^{3/4}\right] -1\right)^{-1}$ with the disk scale length varying with wavelength as $r_s(\lambda)=(\lambda/\lambda_0)^p \, r_s(\lambda_0)$, where $\lambda_0=1736$ \AA\ at the rest frame.
We used $2000\times2000$ magnification maps computed using the inverse polygon mapping algorithm (Mediavilla et al. 2006, Mediavilla et al. 2011a) with 0.5 light-day pixels and $1\,\rm{M_{\sun}}$ stars. All linear sizes can be scaled to a different mass as $(\langle M \rangle/\rm{M_{\sun}})^{1/2}$ and microlensing results are generally insensitive to the mass function (e.g. Wyithe et al. 2000). The maps were then convolved with the disk model using the appropriate size $r_s(\lambda)$ for each wavelength and for each pair of parameters $(r_s, p)$ from a 2D grid of values such that $\ln (r_s^i/\mbox{light-days})=0.3\times i$ for $i=0, \cdots,17$ and $p^j=0.25\times j$ for $j=0,\cdots,9$. For each case we then selected $N=10^8$ random locations in each of the four maps, computed the microlensing magnifications for the different filters, and compared them to the narrowband observations for each epoch. Since this method relies on changes in the microlensing amplitude with wavelength and size but not on its dependence with time, no velocity estimates are necessary. For every 
image $I=(A,B,C,D)$, observed at time $t_j$ and filter $i$, the goodness of the fit is 
\begin{equation}
\label{eq1}
\chi^2(t_j,i)=\sum_I \sum_{J>I}\sigma_{IJ}(t_j,i)^{-2}[\Delta m_I(t_j,i) - \Delta m_J(t_j,i)]^2,
\end{equation}
where
\begin{equation}
\label{eq2}
\Delta m_I(t_j,i)=m_I^{obs}(t_j,i)-\mu_I-\delta \mu_I (t_j,i),
\end{equation}

\noindent
$m_I^{obs}(t_j,i)$ are the data, $\mu_I$ is the macro magnification, $\delta \mu_I (t_j,i)$ is the microlensing magnification,
and $\sigma_{IJ}(t_j,i)$ are the errors as defined in the equation (7) of Kochanek (2004). As described in Mu\~noz et al. (2016), these errors reduce to $\sigma_{IJ}(t_j,i) = 2\sigma(t_j,i)$ if $\sigma_I=\sigma_J(\equiv \sigma)$, and we have chosen to use the average measurement errors of $\sigma=0.08$ mag for weighting all the data.
From this, we estimate the probability density function ${\cal P} (r_s,p)$.

\begin{figure}
\figurenum{2}
\epsscale{0.8}
\plotone{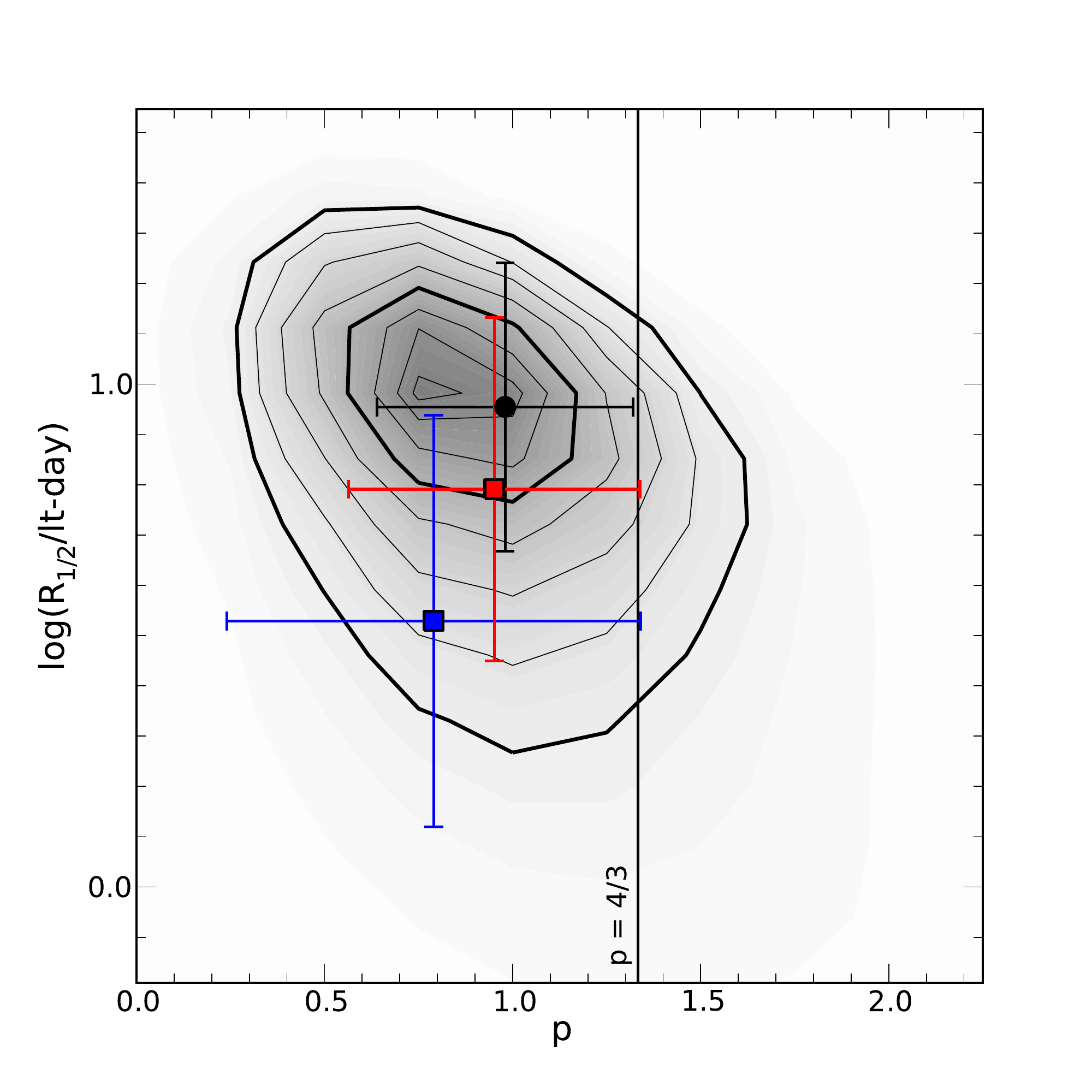}
\caption{\label{pdfh} Joint probability density function ${\cal P}(r_s,p)$ for the half-light radius 
\textbf{$R_{1/2}=2.44 r_s$} (at rest $\lambda_0=1736$ \AA) and the logarithmic slope $p$ ($r_s\propto \lambda^p$) for the disk model using the weighted average of the mid-IR flux ratios from Agol et al. (2000, 2001, 2009), Minezaki et al. (2009), and this work.
The separation between consecutive contours corresponds to 0.25$\sigma$, and the 1$\sigma$ and 
2$\sigma$ contours for one parameter are heavier. The red filled square is the Bayesian estimate for the
expected values of $R_{1/2}$ and $p$ for the averaged mid-IR flux ratios, and the blue filled square is the result of the same calculation using our 2013 mid-IR observations only. The black filled circle corresponds to the measurement by \cite{paper2237} using the mid-IR flux ratios from Minezaki et al. (2009).
All sizes are scaled to a mean stellar mass of $\langle M \rangle= 0.3\,\rm{M_{\sun}}$.
The $p=4/3$ vertical line corresponds to the temperature slope for the standard thin accretion disk model. }
\label{micro}
\end{figure}

Here we use our new mid-IR flux ratios as the intrinsic flux ratios $\mu_{IJ}^{ir}=\mu_I - \mu_J$ instead of those from Minezaki et al. (2009). The results for the expected values of the disk parameters are $r_s=1.40_{-0.85}^{+2.19}\sqrt{\langle M \rangle/0.3\,\rm{M_{\sun}}}$ light-days (equivalent to a half-light radius $R_{1/2}=3.4_{-2.1}^{+5.3}\sqrt{\langle M \rangle/0.3\,\rm{M_{\sun}}}$ light-days) and $p=0.79\pm0.55$, where we have
scaled the results to a mean stellar mass of $\langle M \rangle=0.3\,\rm{M_{\sun}}$.
A logarithmic slope of $p=4/3$ corresponds to a standard thin disk. 
As can be seen in Table~\ref {ratios} and discussed in Section 3, the mid-IR flux ratios reported by different authors and at different epochs
are not mutually consistent given their uncertainties. For comparison to simply using the estimate from our new data, we repeated the calculation using an error-weighted average of all the available mid-IR data
(the ``average" entry in Table~\ref {ratios}). In this case we obtain a scale radius of $r_s=2.5_{-1.4}^{+3.0}\sqrt{\langle M \rangle/0.3\,\rm{M_{\sun}}}$ light-days, a half-light radius of $R_{1/2}=6.2_{-3.4}^{+7.4}\sqrt{\langle M \rangle/0.3\,\rm{M_{\sun}}}$ light-days, and $p=0.95\pm0.39$. Figure~\ref{micro} shows the contours of the probability density function (PDF) using this weighted average along with the results using only our new mid-IR flux ratios, as well as our earlier results from Mu\~noz et al. (2016) using the Minezaki et al. (2009) flux ratios. Despite the differences in the mid-IR flux ratios, all these estimates for $R_{1/2}$ and $p$ are mutually consistent.

The small changes observed in the mid-IR flux ratios over time are likely dominated by systematic errors, but an alternative explanation is that there is mid-IR variability induced by stellar microlensing of the mid-IR emission from the disk \citep{Sluse13}. 
If we assume that this variability is due to microlensing, we can then infer the size of the emitting region using a similar method to
 the one above. For this analysis we generated magnification maps for the four lensed images of the quasar that are $2000\times2000$ pixels with a size of 4 light-days~pixel$^{-1}$ for stars with a mass of $\langle M \rangle = 1\,\rm{M_{\sun}}$. We then convolve them with simple Gaussian models for the source, $I(R) \propto \exp(-R^2/2r_s^2)$, such that $\log_{10} (r_s^i/\mbox{light-days})=1+0.15\times i$ for $i=0, \cdots,19$, and the values of $r_s$ span from 10 to roughly 7000 light-days. Using the weighted average of all observations as an estimate for the baseline with no microlensing and a logarithmic prior, we obtain a Bayesian estimate for the scale radius of the Gaussian source of $r_s=194_{-91}^{+171}\sqrt{\langle M \rangle/0.3\,\rm{M_{\sun}}}$ light-days, which corresponds to a half-light radius of $R_{1/2}=228_{-107}^{+201}\sqrt{\langle M \rangle/0.3\,\rm{M_{\sun}}}$ light-days. 
We then repeated the calculations using the radio flux ratios from Falco et al. (1996) to define the intrinsic flux ratios. Because of the large uncertainties on the radio fluxes, we obtain 
only a lower limit for the size of the mid-IR emission, with $R_{1/2} > 340\,\sqrt{\langle M \rangle/0.3\,\rm{M_{\sun}}}$ light-days.

Since the mid-IR wavelengths correspond to $\lambda \sim$ 4 $\mu$m in the rest frame, the main contribution to the mid-IR emission in this lensed quasar should be dust emission. Dust cannot be closer to the central engine than the point where it would be heated to its evaporation 
temperature. For a simple thermal equilibrium, ignoring Planck factors, this corresponds to a radius of $r_{min} \simeq 730 L_{46}^{1/2} T_{d3}^{-2}$~light-days where the luminosity of the active galactic nucleus (AGN) is $L = 10^{46}\ L_{46}$~erg~s$^{-1}$ and the dust destruction temperature is $T_d = 1000 \,T_{d3}$~K. Agol et al. (2009) estimated
a luminosity of $L = 4\times 10^{46}$~erg~s$^{-1}$ corresponding to $r_{min} \simeq 1500 \,T_{d3}^{-2}$~light-days.
Mid-IR interferometric observations of AGNs point to a torus size of approximately $R_{1/2}\lesssim3$~pc for this luminosity \citep{Burtscher13}. The time scale for microlensing variability of an emissivity region this large would be many decades rather than years, reducing the likelihood that the differences can be due to microlensing (Stalevski\etal\ 2012).
Our default hypothesis, that the apparent ``variability" is really an indication that there are 
systematic errors in the mid-IR fluxes (or their uncertainties), is likely correct, and we should view these estimates for the size of the dusty torus
just as a lower limit with $R_{1/2} \gtrsim 200\,\sqrt{\langle M \rangle/0.3\,\rm{M_{\sun}}}$ light-days.
Alternatively, \cite{Sluse13} suggest that there is still a sufficient contribution from disk emission at these wavelengths 
to produce low levels of microlesning variability, especially in the case of $\kappa_{\ast}/\kappa=1$, which would lead to smaller size estimates than expected from the predicted dust sublimation radius. 

\section{Dark matter substructure}\label{sec5}

Beyond problems in the macro models, the alternate interpretation of differences between the mid-IR flux ratios and smooth models is magnification perturbations due to substructure in the lens. 
In this section we will assume that the mid-IR flux anomalies between our observations and those predicted by our simple smooth SIE+$\gamma$ or NFW+de Vaucouleurs+$\gamma$ models are caused by (dark matter) subhalos orbiting the lens galaxy and acting as ``millilenses''. The goal is to estimate $\beta=b/b_0$, the ratio of their average Einstein radius
$b$ to that of the lens galaxy $b_0$, and their abundance $\alpha$, the fraction of the lensing convergence $\kappa$ that is in the form of subhalos. Since we are using only magnifications, we should not be able to determine $\beta$, but should be
able to constrain $\alpha$.

For each pair $(\alpha, \beta)$ we calculate magnification maps for each of the images of the quasar using the inverse polygonal mapping algorithm, but this time using pseudo-Jaffe density profiles $\rho \propto r^{-2}(r^2 + a^2)^{-1}$ (see Mu\~noz, Kochanek \& Keeton 2001) instead of point masses. We set the scale $a$ as the tidal radius of the subhalo, $a=\sqrt{b\, b_0}$ (Dalal \& Kochanek 2002). We use satellite mass fractions of $\alpha_j=2^{-j}$ for $j=2, \cdots, 8$ and the Einstein radius ratios of $\beta_i=b_0^{-1}(10^{-4}\times 2^i)$ for $i=0, \cdots, 8$. The mass of the individual subhalos spans roughly from $2\times10^4 \,\rm{M_{\sun}}$ to $8\times10^7 \,\rm{M_{\sun}}$. Given the large size expected for the dusty torus (see the discussion in section 4), we calculated magnification maps with a pixel scale of 1000~light-days and a size of $200\times200$ pixels. These regions are still small enough for the millilensing magnification maps associated with each image to be statistically independent. However, when the mass of the millilenses is at the upper end of our range and the abundance is low, only part of one caustic (if any) will be present, and for the smallest subhalos and highest abundances the number of lenses can be so high as to create computational challenges. In the first case, the solution is to generate a larger number of maps to get good statistics, while in the latter case, the size of the map (and/or the area where the lenses are placed, since border effects will be less important when the mass distribution consists of very large numbers of very small subhalos) has to be reduced. 
In any case, our approach assumes an upper limit on the subhalo masses to avoid both correlations between the magnification maps for different images and ray deflections so large that they would require modifications to the macrolens model. The procedure is explained in more detail in Vives-Arias et al. (2016, in preparation).

For each quasar image $I$, the millilensing magnification is 
\begin{equation}
\label{mag}
\Delta \mu_I = m_I - m_0 - \mu_I 
\end{equation}
where $m_0$ is the unknown intrinsic magnitude of the source, $m_I$ is the observed magnitude of image $I$ and $\mu_I$ is the macromodel magnification for that image. If we consider the millilensing magnifications for each of the quasar images as independent events, we can define the probability of observing millilensing magnifications $\Delta \mu_I$ given the parameters $\alpha$ and $\beta$ as
\begin{equation}
\label{sub1}
P(\Delta \mu_I|\alpha,\beta) = \prod_{I=A,B,C,D} P_I(\Delta \mu_I |\alpha,\beta)
\end{equation}
where the $P_I(\Delta \mu_I |\alpha,\beta)$ are the individual PDFs for each image calculated from the magnification maps. If we then substitute equation~\ref {mag} into equation~\ref {sub1} and marginalize over the unknown source magnitude $m_0$, we have
\begin{equation}
\label{sub2}
P_{marg}(m_I-\mu_I |\alpha,\beta) = \int \prod_{I=A,B,C,D} P_I(m_I-m_0-\mu_I|\alpha,\beta)dm_0
\end{equation}
assuming a uniform prior for $m_0$ over the range considered.

\begin{figure}
\figurenum{3}
\epsscale{1}
\plotone{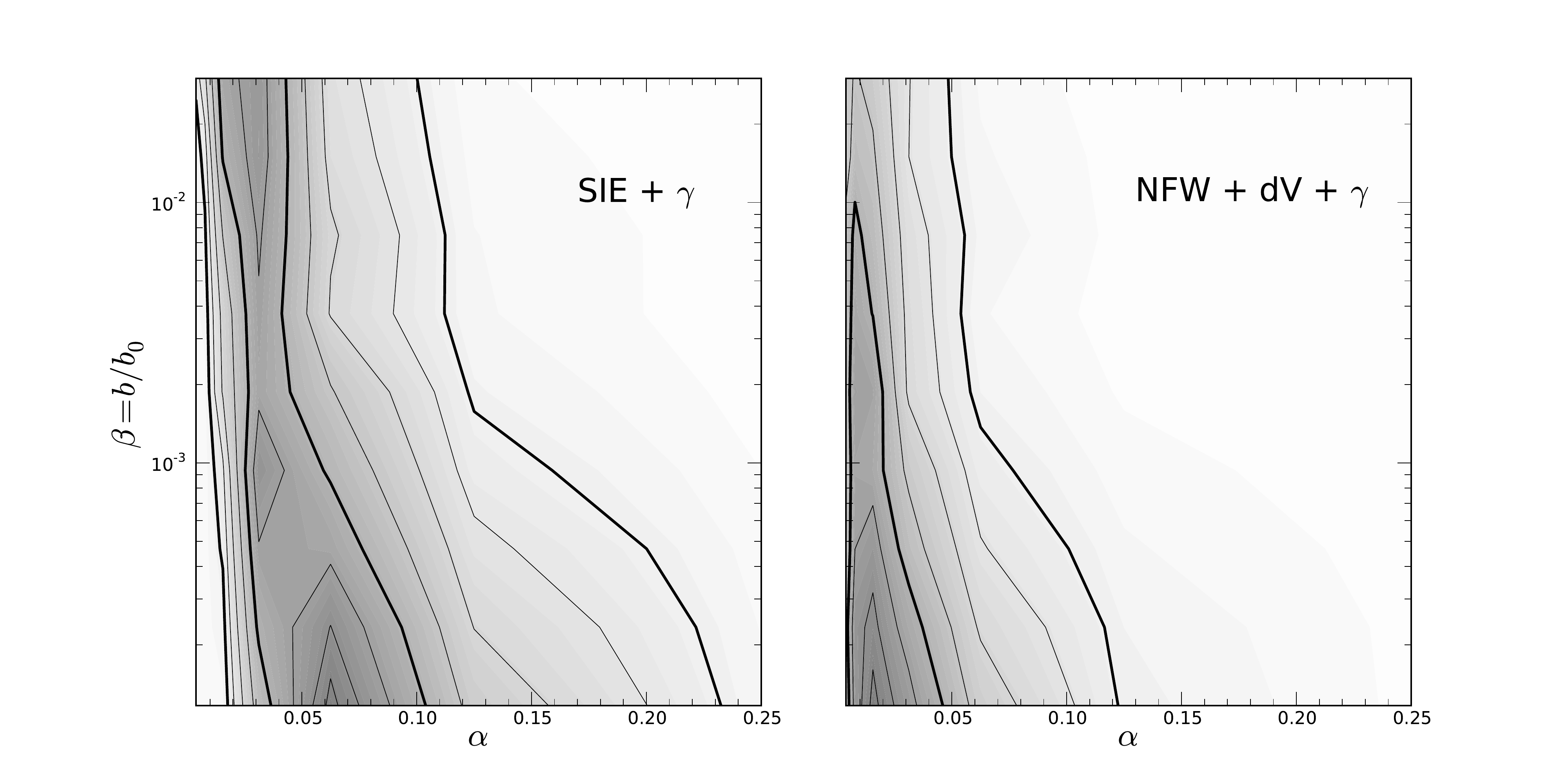}
\caption{Probability densities for a distribution of (dark matter) subhalos around Q2237+0305 in terms of their local mass fraction $\alpha$, and their Einstein radius $b$ expressed as a fraction $\beta=b/b_0$ of the Einstein radius of the SIE profile that best fits the lens galaxy.
The left panel uses the SIE+$\gamma$ model and the right panel uses the NFW+de Vaucouleurs model. The separation between consecutive contours corresponds to 0.25$\sigma$, and the 1$\sigma$ and 2$\sigma$ contours for one parameter are heavier.}
\label{mili}
\end{figure}

Figure~\ref{mili} shows the resulting PDFs for $\alpha$ and $\beta$ corresponding to the SIE+$\gamma$ and NFW+de Vau\-couleurs+$\gamma$ lens models assuming that the true flux ratios are given by the weighted average of all mid-IR observations. We cannot estimate the mass scale of the subhalos $b$; however, their mass fraction $\alpha$ is reasonably well constrained. The SIE+$\gamma$ profile gives an estimate for the abundance of subhalos $\alpha = 0.033_{-0.019}^{+0.046}$ (Figure~\ref{mili}, left). If we repeat the analysis with the prediction of the NFW+de Vaucouleurs+$\gamma$ model, we obtain $\alpha = 0.013_{-0.008}^{+0.019}$ (Figure~\ref{mili}, right). This shows that a small amount of (dark matter) substructure suffices to explain the flux ratio anomalies that smooth macromodels struggle to fit properly.

\section{Discussion and Conclusions}\label{sec6}

We have measured the mid-IR flux ratios at $11.6 \mu$m ($4.3 \mu$m in the rest frame) of the quadruple lens system Q2237+0305 with the CanariCam imager at the GTC. Compared with previous results in the literature, we found
moderately significant differences ($\sim2\sigma$) given the error estimates. Similar differences are seen between the various prior mid-IR flux ratio measurements. 
These differences have little effect on estimates of the properties of the quasar accretion disk. We repeated
our estimates of the size and temperature profile of the disk from \cite{paper2237}. The results are mutually consistent
whether we use the mid-IR flux ratios from Minezaki et al. (2009) that we used in \cite{paper2237}, our new 
mid-IR flux ratios, or the weighted average of all available flux ratios. In particular, we found a disk half-light radius of
$R_{1/2}=6.2_{-3.4}^{+7.4}\sqrt{\langle M \rangle/0.3\,\rm{M_{\sun}}}$ light-days at $\lambda_{rest}=1736$ \AA, and wavelength scale $R\propto \lambda^p$ of $p=0.95\pm0.39$ using the 
weighted average of the flux ratios, where a standard thin-disk model would have $p=4/3$. These results are also consistent with previous estimates based on different approaches to the microlensing calculations (e.g. Poindexter \& Kochanek 2010, Sluse et al. 2011, Mosquera et al. 2013).


The observed variability of the mid-IR flux ratios in different epochs could be due to systematics, but we also considered microlensing by the stars in the lens galaxy as an alternative explanation. 
Under this hypothesis, we obtain an estimated size for the mid-IR emission region assuming a Gaussian source of $R_{1/2}=228_{-107}^{+201}\sqrt{\langle M \rangle/0.3\,\rm{M_{\sun}}}$ light-days. This is smaller than the size expected for the mid-IR emission form a hot dusty torus in the AGN.
This could be due to either underestimated or systematic uncertainties in the mid-IR flux ratios or a residual contribution from microlensing of the more compact disk even at these wavelengths \citep{Sluse13}. It is probably better to regard this estimate as a lower limit.

Finally, assuming that the observed mid-IR flux anomalies with respect to the predictions of simple macromodels are due to (dark matter) substructure, we estimate the mass fraction in satellites that would be needed to reproduce the mid-IR observations. For the flux ratios predicted by an SIE+$\gamma$ model we found $\alpha = 0.033_{-0.019}^{+0.046}$, and for an
NFW+de Vaucouleurs+$\gamma$ model, $\alpha = 0.013_{-0.008}^{+0.019}$. As expected from simply fitting flux ratios, no constraint is found in the mass of the satellites. These results are consistent with CDM predictions (e.g. Zentner \& Bullock 2003) and the observational results obtained by both Dalal \& Kochanek (2002) and Vegetti et al (2014). They also bring down the high estimate obtained by Metcalf et al. (2004) for Q2237+0305 based on the narrow-line flux ratios of this system.

\bigskip

\noindent Acknowledgments:
This research was supported by the Spanish MINECO with the grants
AYA2013-47744-C3-3-P and AYA2013-47744-C3-1-P. J.A.M. is also supported by the Generalitat
Valenciana with the grant PROMETEO/2014/60. C.S.K. is supported by NSF grant AST-1515876. J.J.-V. is supported by the project AYA2014-53506-P financed by the Spanish Ministerio de Econom\'{\i}a y Competividad and by the Fondo Europeo de Desarrollo Regional (FEDER), and by project FQM-108 financed by Junta de Andaluc\'{\i}a. The authors thankfully acknowledge the computer resources, technical expertise, and assistance provided by the ``Centre de C\`alcul de la Universitat de Val\`encia'' through the use of the Lluis Vives and Multivac computing clusters. Based on observations made with the Gran Telescopio Canarias (GTC), installed in the Spanish Observatorio del Roque de los Muchachos of the Instituto de Astrofisica de Canarias, on the island of La Palma.


\clearpage

\clearpage





%

\end{document}